\documentclass[aps,preprint,superscriptaddress,longbibliography,floatfix]{revtex4-1}
\usepackage{graphicx}% Include figure files
\usepackage{color}

\begin{document}
%
%Title of paper
\title{Mott transition and collective charge pinning in electron doped Sr$_{2}$IrO$_4$}
\author{K. Wang}
\affiliation{Department of Quantum Matter Physics, University of Geneva, 24 Quai Ernest-Ansermet, 1211 Geneva 4, Switzerland}
\author{N. Bachar}
\affiliation{Department of Quantum Matter Physics, University of Geneva, 24 Quai Ernest-Ansermet, 1211 Geneva 4, Switzerland}
\author{J. Teyssier}
\affiliation{Department of Quantum Matter Physics, University of Geneva, 24 Quai Ernest-Ansermet, 1211 Geneva 4, Switzerland}
\author{W. Luo}
\affiliation{Department of Quantum Matter Physics, University of Geneva, 24 Quai Ernest-Ansermet, 1211 Geneva 4, Switzerland}
\author{C. W. Rischau}
\affiliation{Department of Quantum Matter Physics, University of Geneva, 24 Quai Ernest-Ansermet, 1211 Geneva 4, Switzerland}
\author{G. Scheerer}
\affiliation{Department of Quantum Matter Physics, University of Geneva, 24 Quai Ernest-Ansermet, 1211 Geneva 4, Switzerland}
\author{A. de la Torre}
\affiliation{Department of Quantum Matter Physics, University of Geneva, 24 Quai Ernest-Ansermet, 1211 Geneva 4, Switzerland}
\affiliation{Institute for Quantum Information and Matter, California Institute of Technology, Pasadena, California 91125, USA}
\author{R. S. Perry}
\affiliation{London Centre for Nanotechnology and UCL Centre for Materials Discovery, University College London, London WC1E 
6BT, United Kingdom}
\author{F. Baumberger}
\affiliation{Department of Quantum Matter Physics, University of Geneva, 24 Quai Ernest-Ansermet, 1211 Geneva 4, Switzerland}
\author{D. van der Marel}
\email[]{dirk.vandermarel@unige.ch}
\affiliation{Department of Quantum Matter Physics, University of Geneva, 24 Quai Ernest-Ansermet, 1211 Geneva 4, Switzerland}
\date{\today}
\begin{abstract}
We studied the in-plane dynamic and static charge conductivity of electron doped Sr$_{2}$IrO$_4$ using optical spectroscopy and DC transport measurements. The optical conductivity indicates that the pristine material is an indirect semiconductor with a direct Mott-gap of 0.55~eV. Upon substitution of 2$\%$ La per formula unit the Mott-gap is suppressed except in a small fraction of the material  (15\%) where the gap survives, and overall the material remains insulating. Instead of a zero energy mode (or Drude peak) we observe a soft collective mode (SCM) with a broad maximum at $40$ meV. Doping to 10$\%$ increases the strength of the SCM, and a zero-energy mode occurs together with metallic DC conductivity. Further increase of the La substitution doesn't change the spectral weight integral up to 3 eV. It does however result in a transfer of the SCM spectral weight to the zero-energy mode, with a corresponding reduction of the DC resistivity for all temperatures from 4 to 300 K. The presence of a zero-energy mode signals that at least part of the Fermi surface remains ungapped at low temperatures, whereas the {\color{black} SCM} appears to be caused by pinning a collective frozen state involving part of the doped electrons.
\end{abstract}
% 
%\keywords{}
%
\maketitle
\section{Introduction}
Doped Mott insulators have been found to exhibit a rich spectrum of remarkable physical phenomena, including metal-insulator transitions, charge and spin ordering, stripe order, orbital currents, high-T$_c$ superconductivity, and the pseudo-gap phenomenon~\cite{keimer2015}. 
Sr$_2$IrO$_4$ has a quasi two-dimensional structure of corner sharing IrO$_6$ octahedra. Due to tilting of the octahedra there are two equivalent Ir superlattices organized in a $\sqrt{2}\times\sqrt{2}$ superstructure in the plane.  
The combination of tetragonal crystal field and spin-orbit interaction makes that for each of the Ir atoms two of the $5d$ bands are well below the Fermi energy, $E_F$, and are therefore fully occupied, one band cuts the Fermi surface and is half full, and the two remaining $5d$ bands are far above $E_F$ and therefore empty. The on-site Coulomb energy further splits the half-filled band in a filled lower Hubbard band (LHB) and an empty upper Hubbard band (UHB). Doping this Hubbard insulator results in a strongly correlated metal. It has been suggested that this material can be turned into a high T$_c$ superconductor~\cite{wang2011}. 
{\color{black}While anti-ferromagnetism and spin density wave order have been reported for different doping concentrations~\cite{chen2015,gretarsson2016,chen2018}}, up to date transport and magnetization data have not revealed superconductivity~\cite{chen2015,delatorre2015}. 
A recent angle resolved photoemission (ARPES) study found an anisotropic pseudogap in Sr$_{2-y}$La$_y$IrO$_4$~\cite{delatorre2015}. ARPES~\cite{kim2016} and STM~\cite{yan2015} experiments on K-covered Sr$_2$IrO$_4$ further reported evidence for a d-wave gap closing at 30-50~K. In addition a degeneracy splitting of the bands near $(\pi,0)$ was found in ARPES experiments~\cite{delatorre2015} and a hidden order parameter has been claimed based on observations with optical second harmonic generation~\cite{zhao2015}. These observations have lead to different mutually exclusive speculations as to the nature of this state of matter, in particular superconducting fluctuations~\cite{kim2016} and a d-wave pseudospin-current ordered state~\cite{zhou2017}. ARPES and STM probe the single electron spectral function. Information on the {\em collective} current response requires on the other hand measurements of the optical conductivity. 

Here we use optical spectroscopy from 12.5~meV to 4~eV and DC transport experiments of pristine and doped Sr$_2$IrO$_4$ to measure the doping evolution of the free carrier density and the optical spectra. 
{\color{black} We report the following new results:  
(i) We demonstrate the appearance of a MIR mode at 0.2~eV for all dopings, the intensity of which tracks the charge carrier concentration. This feature is common with other doped Mott systems such as the cuprates.
(ii) We demonstrate that a rapid collapse of the Mott gap is obtained with doping. However, an important part of the doped carrier response shows up in a soft collective mode at finite frequencies at the detriment of the spectral weight of the Drude peak even for dopings as high as y=0.1 and to a lesser extent for y=0.18.
(iii) We present theoretical calculations the optical conductivity of the Mott-insulating parent compound, as well as the doped material, using the self-consistent Hartree-Fock approximation. The calculations reproduce the single-particle bandstructure as measured with ARPES, and the doping dependence of the Drude spectral weight. However, the effects of doping at finite energy are not fully captured by this approach, thus motivating the development of alternative theories for this class of materials.  
(iv) We determine the quasi-linear doping dependence of the spectral weight of the zero-energy mode (the Drude peak), from which we obtain the kinetic energy $K^*(y)$ of the renormalized charge carriers, and show that $K^*(y)$ is approximately 5 times smaller than in the hole-doped cuprates.
(v) Our observations point toward a scenario where the collective charge sector is composed of two components, one associated to ungapped fermions, and the other to {\color{black} a frozen correlated state of the electrons (for example a charge density wave)} pinned by disorder associated with the donor states. }
 
\section{Methods and Results}
\begin{figure}[t!!]
\includegraphics[width=\columnwidth]{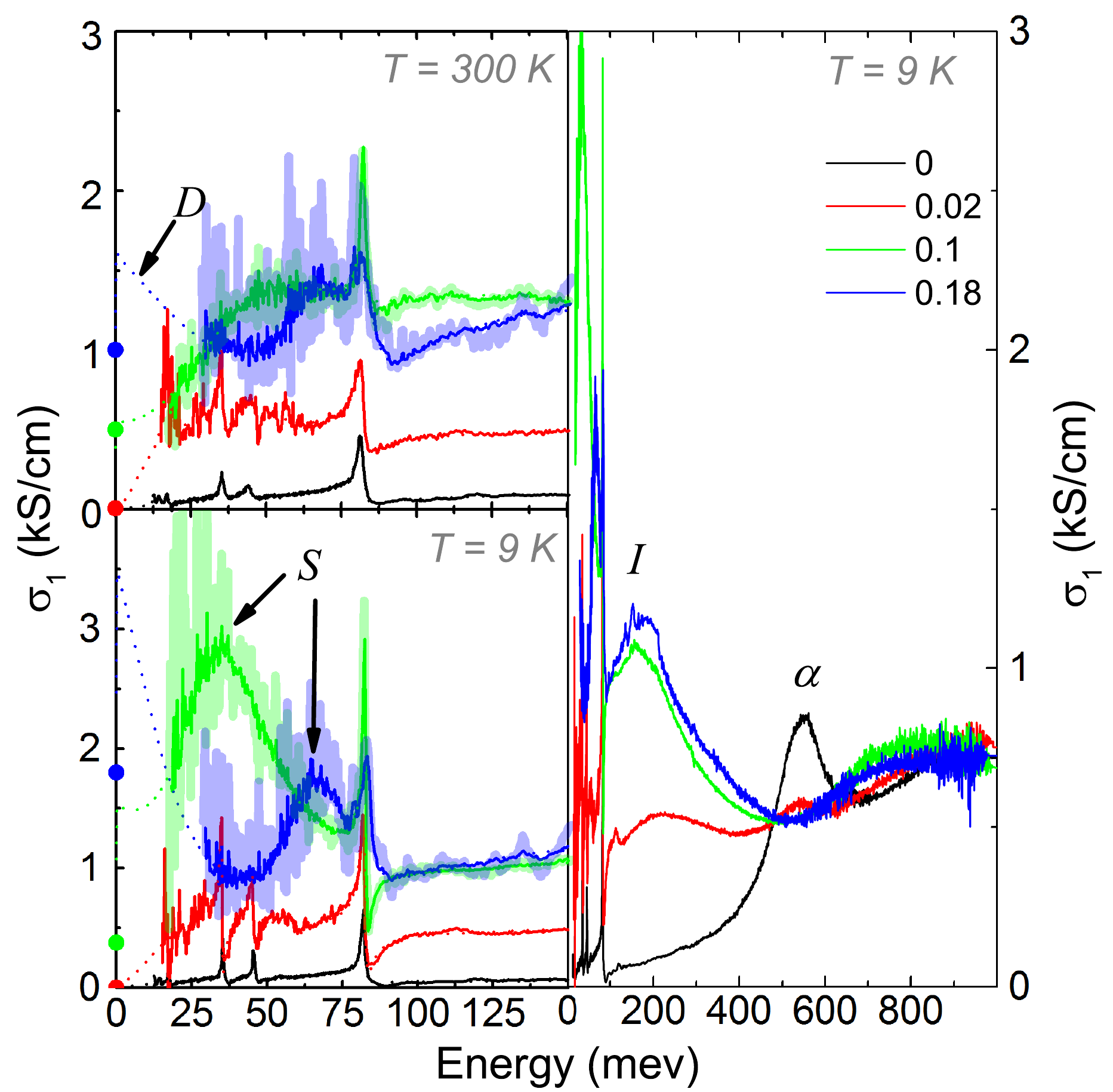}
\caption{\label{fig:sigma_IR} Left: doping dependence of the optical conductivity from 0 to 0.15~eV at 300~K and 9~K. The dashed curves represent extrapolations of the optical conductivity obtained from fitting reflectivity $R(\omega)$ with a Drude-Lorentz parametrization. The shaded areas correspond to the noise level of the original reflectivity data (see Fig.~\ref{fig:reflectivity}). The dark lines are the result of binning the data in intervals of 0.3~meV. Full symbols correspond to the conductivity values measured in DC transport (see Fig.~\ref{fig:resistivity}). 
Right: doping dependence in the range of 0 to 1~eV at 9~K.}
\end{figure}
Single crystals of Sr$_{2-y}$La$_y$IrO$_4$ were grown as described in Appendix~\ref{app:experiments}.  {{\color{black} Stoichiometry and doping concentrations are reported in Table  \ref{table:doping}}.
The complex dielectric function and optical conductivity were determined in the range of 12.5  meV to 4 eV by combining reflectivity and ellipsometry methods as described in Appendix~\ref{app:experiments}. 
The resulting low energy optical conductivity for different temperatures and dopings is displayed in Fig.~\ref{fig:sigma_IR}.
For the doped samples we observe a zero-energy mode ($D$). The zero energy conductivity corresponds to the inverse of the DC resistivity which has been measured using standard transport methods and is shown in Fig.~\ref{fig:resistivity}. The DC resistivity was measured of samples of the same composition, grown under identical conditions. The $y=0.1$ and $y=0.18$ samples approaches a linear temperature dependence at high temperature, and has a resistivity upturn at low temperatures (see inset of Fig.~\ref{fig:resistivity}), whereas the resistivity of the $y=0.0$ and $y=0.02$ samples has a negative slope at all temperatures indicating insulating or at best bad metal behavior. The fact that the DC conductivities (full symbols in Fig.~\ref{fig:sigma_IR}) and the peak maxima of the zero-energy mode do not coincide, indicates that the spectrum below the measured range (20 meV for this sample) is not fully described by the Drude-Lorentz model employed here. We emphasize however, that despite the lack of detail below 20 meV, the spectral weight in this range can be obtained accurately by analyzing the real and imaginary dielectric function at higher frequencies \cite{kuzmenko2007}. 
In the range between 0 and 100 meV we observe optical phonon excitations. These modes are most prominent in the undoped compound (see Table~\ref{table:phonons}). Upon doping all modes exhibit Fano-type asymmetries. With the exception of the 82 meV mode, the phonons disappear against the electronic background in the samples with doping of $y=0.1$ and $y=0.18$, signaling an effective channel for charge screening at high doping.
Upon doping all modes exhibit Fano-type asymmetries. With the exception of the 82 meV mode, the phonons disappear against the electronic background in the samples with doping of $y=0.1$ and $y=0.18$, signaling an effective channel for charge screening at high doping.
Upon doping $y=0.02$ a SCM appears with a maximum at 40 meV (labeled $S$), which sharpens and gains spectral weight for $y=0.1$ and collapses into the Drude peak for $y=0.18$. For the same doping a SCM appears at 60 meV. Since the spectral weight in these modes is far too high for phonons, they have to correspond to electronic collective modes. These modes strongly couple to the optical phonons, which is strikingly demonstrated by the E$_u$(4), E$_u$(5) and  E$_u$(6) modes (see Table~\ref{table:phonons}), each overlapping with the energy of the {\color{black} SCM}. 
{\color{black}Fano-type asymmetry has also been observed for the Raman active phonons in the doped material \cite{Gretarsson2017}. For the y=0.02 sample it can not be excluded that this asymmetry may -at least in part- have to do with electronic heterogeneity that we will discuss in following section. However, since phonon asymmetry is a commonly observed and well understood consequence of electron-phonon coupling~\cite{rice1977,damascelli1997}, this interpretation likely applies to all samples. The absence of the E$_u$(4) and E$_u$(5)  phonons from the $y=0.1$ and $y=0.18$ spectra is probably due to strong mixing with the {\color{black} SCM}.}
We notice that some asymmetry is also present for the undoped compound, along with a weak background optical conductivity that gradually rises to the maximum at 550 meV (peak $\alpha$ that we will discuss below). 
For all doped samples a prominent MIR peak is centered at about 0.2~eV (peak $I$) for $y=0.1$, which redshifts upon increasing the doping. 
This feature is commonly observed in doped Mott insulators, in particular the high-$T_c$ cuprates, and is associated to the dressing of the charge carriers by dynamical degrees of freedom of the system~\cite{hwang2004,heumen2009}. The exact nature of these degrees of freedom is difficult to determine unambiguously; theoretical interpretations include phonons, fluctuations of spin, charge and loop-current, and combinations thereof~\cite{varma1987,kane1989,rice1989,tachiki1988,grueninger1999}. 

\begin{figure}[t!!]
\includegraphics[width=\columnwidth]{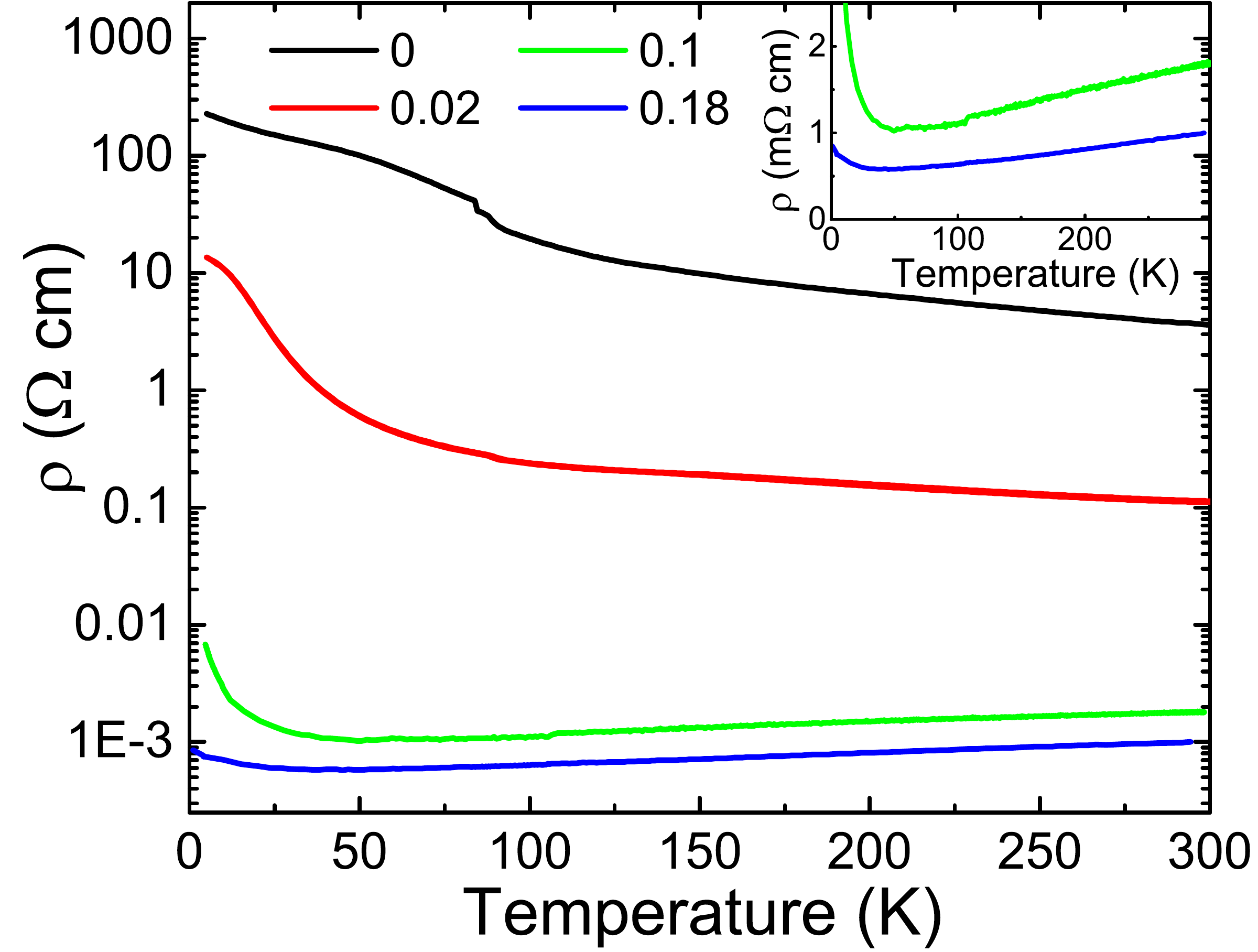}
\caption{\label{fig:resistivity} Temperature dependence of the resistivity for different doping compositions.}
\end{figure}
\begin{figure}[t!!]
\includegraphics[width=\columnwidth,angle=0]{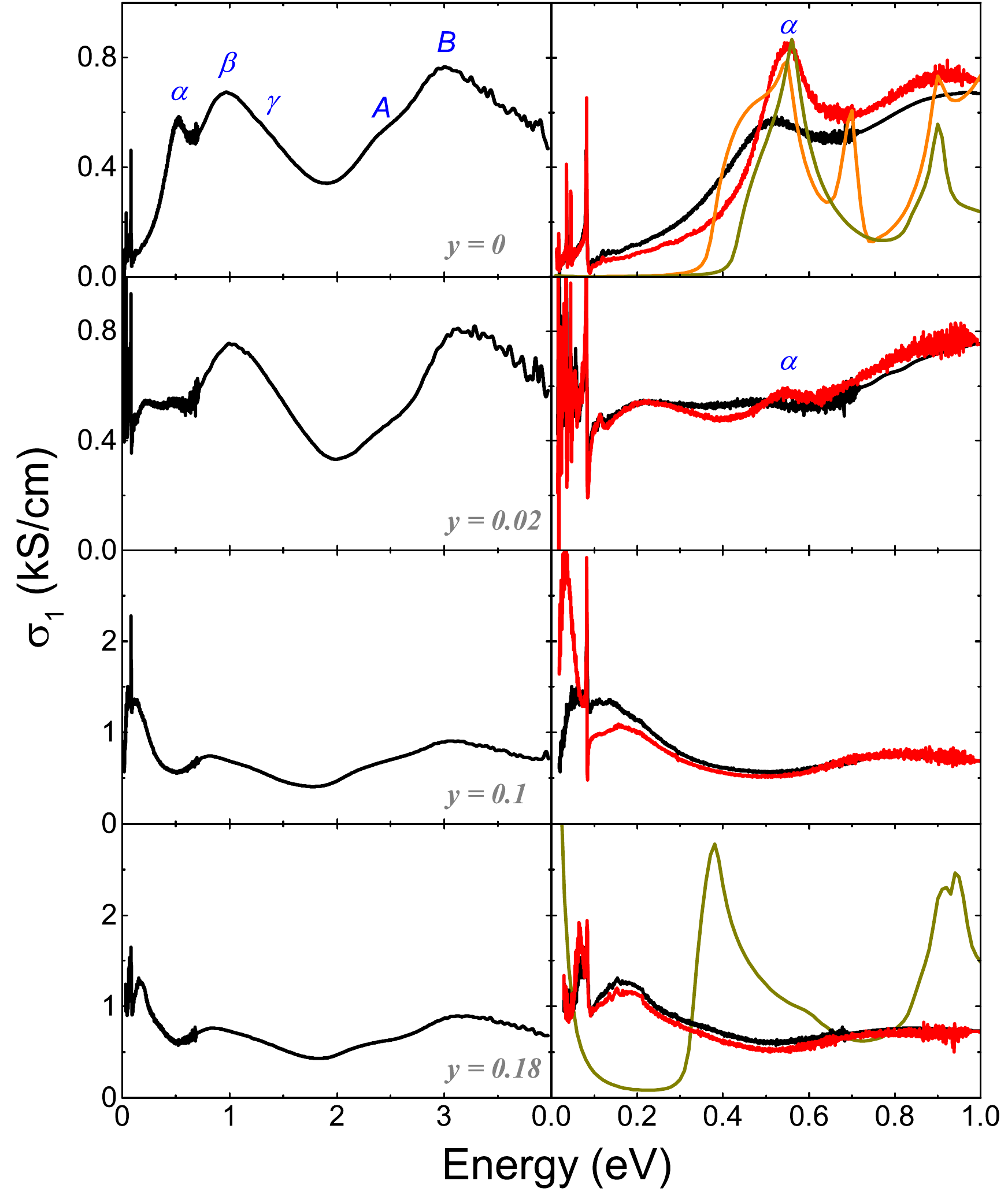}
\caption{\label{fig:sigma_broad} Left: doping dependence of optical conductivity in the whole measured range.
Right: temperature and doping dependence of the real part of optical conductivity Below 1eV. The spectral features are labeled according to Ref.~\onlinecite{moon2009}. {\color{black}Tight-binding calculations for the undoped material at 30 K using the parameters A (same parameters as Ref.~\onlinecite{delatorre2015}) and B (best fit to the optical data) of Table~\ref{table:theory} of Appendix~\ref{app:theory} are shown as orange (A) and dark-khaki (B, best fit). The result of model A (B) has been scaled down by a factor 5 (2.5) to match the vertical range of the data shown. The dark-khaki curve for $y=0.18$ was calculated with the same band parameters as the dark-khaki curve of the top panel (parameter set B) but the self-consistent solution of the bandstructure gives a smaller Hubbard gap due to the electron doping.}}
\end{figure}

To identify the structures in the optical spectra and obtain the parameters for characterizing the electronic structure (Table~\ref{table:theory}) we compare the experimental data with a tight-binding calculation of the optical conductivity of the undoped material (Fig.~\ref{fig:sigma_broad}, see Appendix~\ref{app:theory} for details of the calculation). The orange  curve (model A) corresponds to the parameters used in Ref.~\onlinecite{delatorre2015} to fit the ARPES data. To obtain a better match of peak $\alpha$ we repeated the calculation with a different set of parameters (dark-khaki curve, model B) taking into account the exchange interaction within the Ir-$5d$ shell~\cite{marel1988}. Peak $\alpha$ at 0.55~eV corresponds to the transition from the lower to the upper Hubbard $j=1/2$ band (LHB to UHB) and peak $\beta$ at 1~eV to transitions from the fully occupied $j=3/2$ band to the UHB, which confirms the generally agreed assignment of these peaks~\cite{kim2008,moon2009,sohn2014,propper2016,seo2017,propper2016}.

\section{Discussion of the Results}
The influence of raising the temperature is to deplete the intensity of peak $\alpha$. At least two factors contribute to the observed temperature dependence: loss of short range anti-ferromagnetic correlations~\cite{aichhorn2002}, and change of the Ir-O-Ir bond angles~\cite{moon2009}.  Infrared near-field images (see Appendix~\ref{app:experiments}) of the $y=0.02$ sample clearly indicate insulating islands of various length scales embedded in a metallic bath. STM studies of the doped material have indicated nanoscale regions where the Mott-gap is fully intact, coexisting with metallic regions~\cite{battisti2017}. The presence of the $\alpha$ feature is a natural consequence, which indeed we observe for the $y=0.02$ doped sample (see Fig.~\ref{fig:sigma_broad}). The observed oscillator strength in Fig.~\ref{fig:sigma_IR} corresponds to about 15\% of insulating inclusions.  
{\color{black} In a recent study Seo {\em et al.}~\cite{seo2017} measured the optical spectra of crystals of Sr$_{2-y}$La$_{y}$IrO$_4$ and obtained optical spectra for $y\sim 0.13$ very similar to the $y=0.02$ data of the present study, and much less spectral weight than for the $y=0.1$ sample reported here in the region below 0.5 eV. We speculate that these differences may arise from differences in effective electron doping associated with the oxygen stoichiometry. In particular if there is excess oxygen in the samples, part of the electrons donated by the lanthanum atoms would become trapped by oxygen acceptor states. The data for $y=0.1$ and $y=0.18$ shown in Fig.~\ref{fig:sigma_IR} represent in this respect a considerably higher charge carrier density than recently reported results.} The doping dependent suppression of $\alpha$ has also been observed in epitaxial 
%Sr$_{2-x}$La$_x$IrO$_4$, Sr$_2$Ir$_{1-x}$Rh$_x$O$_4$ and Sr$_2$Ir$_{1-x}$Ru$_x$O$_4$ 
thin films~\cite{lee2012}. 
% Seo's doping levels would than correspond to $y=0.006$ and $y=0.02$.   
The disappearance of peak $\alpha$ for $y>0.02$ signals the collapse of the Mott gap. In comparison, in the hole-doped cuprates the 2 eV charge transfer gap vanishes at a much higher carrier concentration of 0.1 holes per Cu atom~\cite{uchida1991}.  

\begin{figure}[t!!]
\includegraphics[width=\columnwidth]{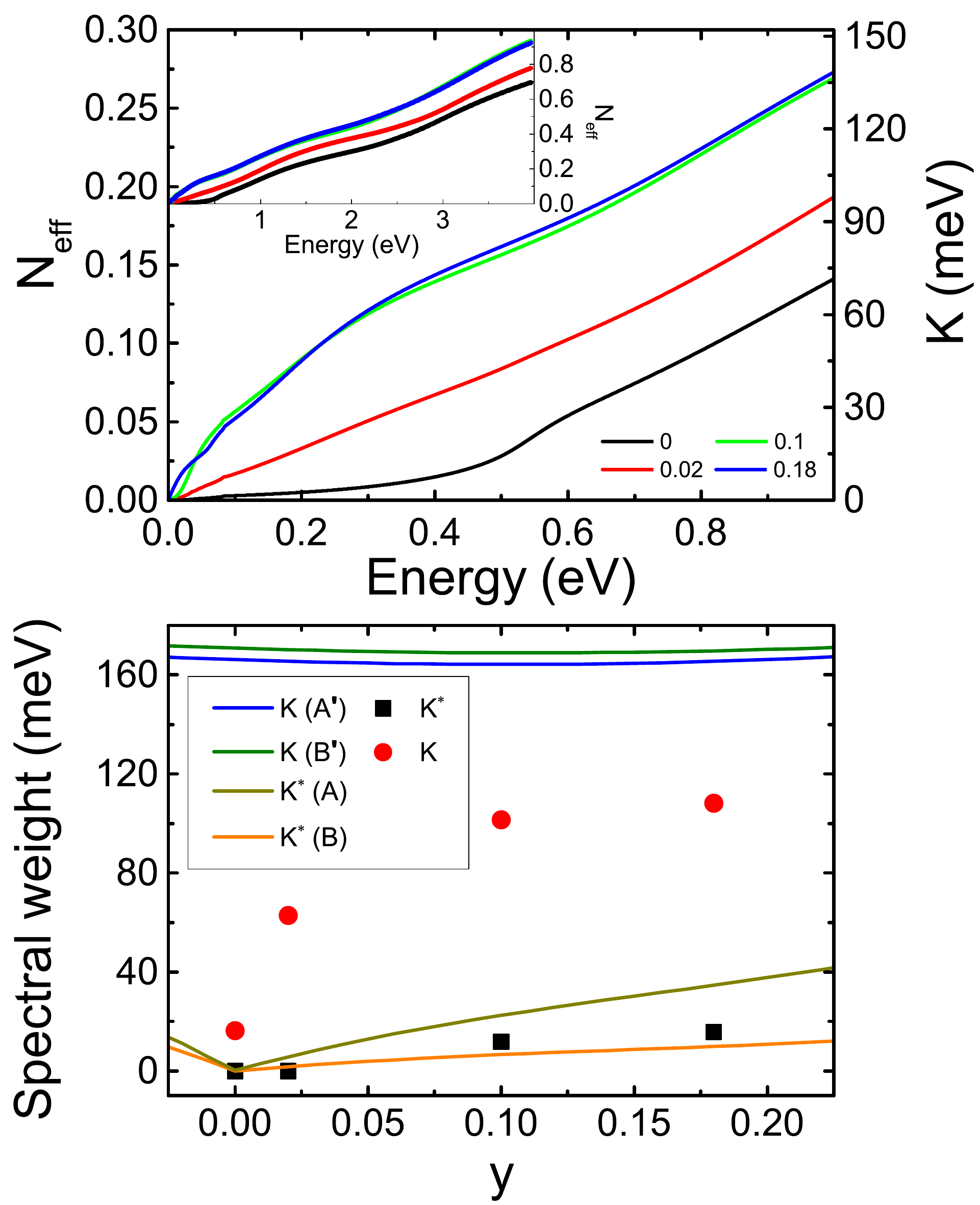}
\caption{\label{fig:spectral_weight} Top: Effective electron number per Ir atom as a function of energy for different dopings. Bottom: Doping dependence of the coherent ($K^*$, black squares) and total ($K$, red circles) free carrier spectral weight. Calculated values are shown for the band parameters reported in Table~\ref{table:theory}. 
}
\end{figure}
For the doping dependence of the low energy spectral weight we use the parameter 
\begin{equation}
K=\frac{d_c}{4 \pi e^2}\sum_j (\hbar\omega_{p,j})^2. 
\end{equation}
Here $d_c$ is the interlayer spacing ({\em i.e.} $d_c=c/4=0.645$ nm), and the plasma frequencies $\omega_{p,j}$ are obtained from the Drude-Lorentz decomposition as detailed in Appendix~\ref{app:experiments} (Fig.~\ref{fig:DL}). In the following discussion we will use the data obtained at 9~K.
The spectral weight of the zero-energy mode, $K^{\ast}$, is obtained by restricting aforementioned oscillator sum to the Drude ($j=0$) peak. This quantity, having units of energy, corresponds to $\epsilon_F/\pi$ for a 2D free electron gas and to $-\epsilon_{kin}/2$ for a single tight-binding band in 2D. The doping dependence displayed in Fig.~\ref{fig:spectral_weight} (bottom panel) follows approximately the linear relation $K^{\ast}\approx K_0 y$ where $y$ is the electron count per Ir atom, as expected for doping the Mott-insulating phase. In the present material $K_0\approx 100$~meV.  In comparison, for the hole doped cuprates $K_0\approx 500$~meV~\cite{mirzaei2013}, indicating 5 times lower spectral weight per quasiparticle than in the cuprates.  In the same figure we compare this to the spectral weight expected from band-structure calculations. The calculated doping dependence without Hubbard interaction (medium-blue and forest-green curves) is very far from our experimental data. In fact these data approximate more closely the behavior expected for a doped Mott-insulator using the Hartree Fock approximation (dark-khaki and orange curves). Despite this reasonably good match these calculations did not reproduce the rapid collapse of the Mott gap at low doping that we observe experimentally, as illustrated in the lower right panel of Fig.~\ref{fig:sigma_broad}.

We now turn to the combined {\em free and bound} intra-band spectral weight obtained by restricting the expression for $K$ to the oscillators below 0.5~eV. 
This spectral weight (Fig.~\ref{fig:spectral_weight}, bottom panel) shows a steep rise from the parent compound to the $y=0.02$ doping, which corresponds to a rapid transfer of high energy to free charge spectral weight, followed by a plateau. Such a behavior is also observed in the cuprates~\cite{uchida1991,eskes1991}. 
The effective electron number can be calculated from the optical conductivity using the relation:
\begin{eqnarray}
{N_{eff}(\omega)} =&&\frac{2 m_{e} V_u}{{\pi}e^2}\int_{0}^{\omega}\sigma_1(\omega^{'})d\omega'
\end{eqnarray}
where $m_{e}$ is the free electron mass and $V_u=97.1~\mbox{\AA}^3$ is the volume of one formula unit, and
$N_{eff}(\infty)$ corresponds to the total number of electrons per formula unit. 
Limiting the  integral to the zero-energy mode provides $K^{\ast}m_eV_u/(d_c\hbar^2)$.  
$N_{eff}(\omega)$ at $T=9$K is displayed in Fig.~\ref{fig:spectral_weight}. The change of slope of the parent compound at 0.55~eV clearly shows the onset of the Mott gap. For the two highest dopings, $N_{eff}(\omega)$ rises steeply from zero energy due to the zero energy mode. 
For all samples the contribution of optical phonons is negligible as compared to the electronic spectral weight.
The effect of doping is to deposit spectral weight below 0.55~eV, while the spectral weight from 0.55 to 4 eV is only weakly affected. 
Doping with $y=0.02$, $y=0.10$ and $y=0.18$ electrons adds amounts $\Delta N_{eff}=0.045, 0.125$ and $0.125$, respectively, in the range below 0.55~eV. Above 0.55~eV the slope of $N_{eff}(\omega)$ is approximately the same for all doping levels, indicating that the effect of doping only weakly affects the optical spectra from 0.55 to 4~eV. Hence we see that for $y=0.02$ and $y=0.10$ {\em more} low energy spectral weight is added than the number of electrons doped. However, the $f$-sum rule implies that for $\omega \rightarrow \infty$ the distance between the $N_{eff}$ curves corresponds to the difference in electron count. The implication is that the extra spectral weight below 0.55~eV is transferred from energies above 4 eV, an effect generally understood to result from strong electron correlations. Indeed the effect of switching on $U$ and $J$ is to transfer spectral weight from the zero-energy mode to higher energies. Integrating the optical conductivity to energies well above the Hubbard $U$ should recover most of the transferred spectral weight. We therefore expect $K$ and $N_{eff}(\omega\sim U)$ to exhibit the doping trend of the band-model with $U=J=0$, which is indeed almost doping independent as shown by the forest-blue and medium-green curves in the lower panel of Fig.~\ref{fig:spectral_weight}. 
  
{\bf Implications for the state of matter:} 
The midinfrared feature $I$ at 0.2 eV (Fig.~\ref{fig:sigma_IR}) is present for all doping concentrations, and its' intensity tracks the doping concentration. Similar features and doping dependence are common in doped Mott insulators (for a summary see section IIIC of Ref.~\onlinecite{basov2011}) and signal the incoherent side-bands of the Drude peak due to coupling of the conduction electrons to dynamical degrees of freedom such as phonons, magnons and combinations thereof. {\color{black}In the cuprates the mid-infrared band has been associated to the pseudo-gap observed with other techniques~\cite{basov2011}. Recent dynamical mean field calculations of doped Sr$_2$IrO$_4$ show that the gap at the ($\pi$,0) point persists at finite doping as a finite-size -but much smaller- pseudogap.\cite{Moutenet2018}} Seeing a pronounced feature like this is typically associated to the regime of low density and strong coupling, resulting in polaronic charge carriers due to coupling to vibrational degrees of freedom. The resulting charge carrier effective masses are typically 2 to 4 times the bare band mass. The possibility of a polaronic nature of the charge carriers in Sr$_2$IrO$_4$ has been previously proposed on the basis of ARPES~\cite{king2013} and optics~\cite{sohn2014}.
The doping dependence of the SCM shows a similar wiping out at high carier concentrations as observed in Nd$_{2-x}$Ce$_x$CuO$_4$~\cite{lupi1999}. 
An interesting explanation of this doping dependence~\cite{lorenzana2001} interprets the {\color{black} SCM observed in  this electron doped cuprate as an} internal mode of the polaron, which should become unstable for large concentration due to dipole-dipole interactions. This in turn leads to a rapid doping dependent softening of this internal mode. This interpretation finds additional support from the observation of a strong Fano-asymmetry of the phonons overlapping with peak $S$ in our measurements. Since both electron-phonon coupling and Mott physics appear to be important in the iridates, and moreover these materials are doped by chemical substitution of donor atoms, it is of interest to consider the effects of doping a Mott insulating state by chemical substitution. In a recent theoretical discussion of R$_{1-x}$Ca$_x$VO$_3$  the effects of disordered charged defects on the spectral function in the Mott insulating regime~\cite{avella2017} were shown to trigger small spin-orbital polarons, with their internal kinetic energy responsible for the opening of the soft defect states gap  inside the Mott gap. 
Breaking of translation invariance is a natural consequence of  the disorder potential due to the randomly substituted La$^{3+}$ donor atoms. The fact that the {\color{black} SCM} has a finite energy requires pinning of the translational motion of the collective charge motion, without which the SCM would show up as a zero-energy mode. 

\begin{figure}[t!!]
\includegraphics[width=\columnwidth]{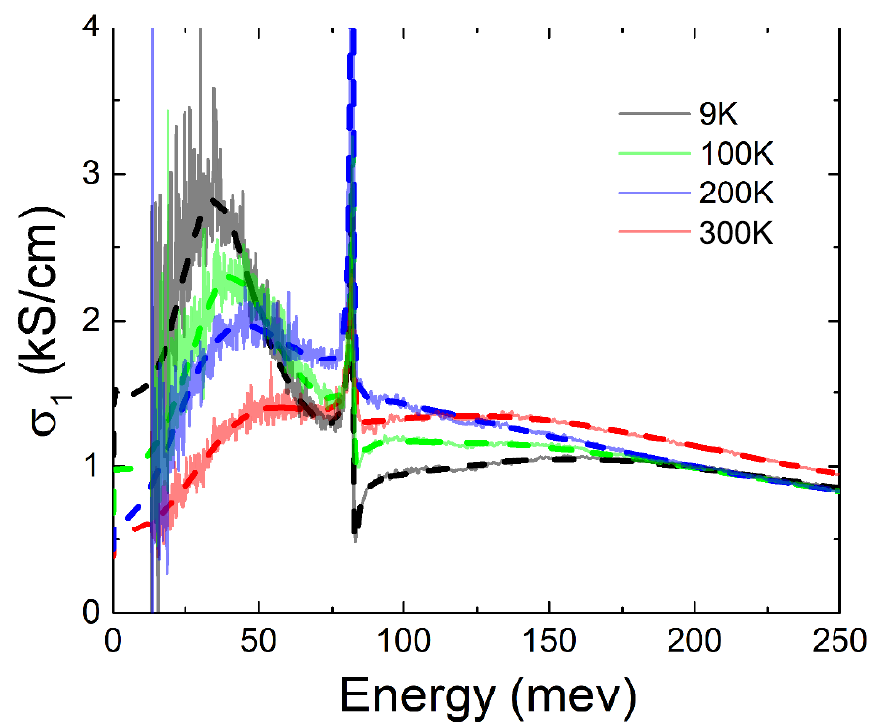}
\caption{\label{fig:T-dependence} \color{black}Zoom of the optical conductivity in the SCM region of sample $y=0.1$. 
}
\end{figure}
All these arguments require that the nature of the charge response is manifestly collective, {\em i.e.} it is necessary to consider the motion of the electronic liquid as a whole as opposed to that of single electrons. 
In a number of recent papers a hydrodynamic instead of corpuscular approach has been explored for the transport \cite{muller2009} and optical~\cite{forcella2014} properties of strongly interacting matter. Recently a theoretical description of the collective hydrodynamic motion of an incommensurate charge density wave state lead to the prediction of optical spectra with a SCM very similar to those shown in Fig.~\ref{fig:sigma_IR}, together with T-linear DC resistivity at high enough temperature. The peak structure in the optical spectra derives in this case from finite energy oscillations of a {\color{black} charge density wave phason} pinned by broken translation invariance of the system~\cite{delacretaz2017a,delacretaz2017b}. Experimentally for the $y=0.1$ and $y=0.18$ samples the pinned collective state appears shunted by a finite DC resistivity, preventing the resistivity from diverging for $T\rightarrow 0$, which is also in good agreement with the model of Ref.~\onlinecite{delacretaz2017a,delacretaz2017b}. It is possible that on a microscopic level this metallic conducting component is characterized by fermions with a Fermi surface of which a finite fraction remains ungapped. 

{\color{black} If we now take a look at the temperature dependence of the SCM, shown in Fig.~\ref{fig:T-dependence} for the $y=0.1$ sample where the SCM is the most clearly manifested, we notice two important trends in the temperature dependence: Increasing temperature causes both a blueshift and a loss of spectral weight of the SCM. This lost SCM spectral weight is transferred to the spectral range above 0.1 eV and almost fully recovered below 0.25 eV. Since Lorenzana did not explicitly work out the temperature dependence~\cite{lorenzana2001}, a direct comparison to Fig.~\ref{fig:T-dependence} can not be made.

In the context of the model of Delacretaz {\em et al.} the strongest temperature dependent blueshift is expected in a quantum critical state of matter~\cite{delacretaz2017a,delacretaz2017b}, in which case $\omega_0\propto k_B T /\hbar$. Our experimental temperature dependence has the same sign, but for $T\rightarrow 0$ the SCM position saturates at 35 meV instead of converging to zero. This saturation at low temperatures would imply that quantum criticality is not realized for the doping values in this study. The emerging picture in the context of the model of Delacretaz {\em et al.} is that of a state of matter characterized by two components existing in parallel: (i) A charge density wave which is pinned to the disorder potential, but can excited at finite energy giving rise to a SCM.  (ii) An interacting electron liquid which is partially ungapped at the Fermi surface. 

On the basis our present data we cannot fully rule out the polaronic interpretation of Lorenzana~\cite{lorenzana2001} or the bad metal interpretation of  Delacretaz {\em et al.}~\cite{delacretaz2017a,delacretaz2017b}. Moreover, both electron-electron interactions\cite{heumen2009} and electron-phonon interactions\cite{mechelen2008} give rise to significant mass-renormalization in low doped transition metals, making it difficult to separate these effects. On the other hand there is an important qualitative difference: The polaronic interpretation~\cite{lorenzana2001} does not require disorder for the SCM to appear at non-zero frequencies, whereas in the work Delacretaz {\em et al.} weak disorder is a requirement for the CDW to appear at non-zero energy. Since in practice weak disorder in materials is a sensitive parameter of preparation conditions, the latter interpretation would then imply considerable  sample-to-sample differences of the SCM. A systematic study of this question will become possible when highly doped Sr$_{2}$IrO$_4$ crystals can be routinely produced.
}

\section{Conclusions}
We observed in the optical conductivity of doped Sr$_{2-y}$La$_y$IrO$_4$ a rapid erosion of the 0.55~eV Mott gap when the material is electron doped by chemical substitution of La on the Sr site. Doping introduces various features below 0.55~eV, in particular a zero-energy mode, an {\color{black} SCM} in the range 35-60~meV and a mid-infrared band at 0.2~eV. The doping evolution of these features indicates that the material remains strongly correlated for all doping values studied up to the maximum doping obtained, $y=0.18$. We also measure a low temperature upturn in the DC resistivity, even at this high doping, whereas the high temperature dependence is approximately  $T$-linear. The 0.2~eV peak is similar to many other doped Mott-insulators, which for this reason we attribute to strong coupling of the electrons to vibrational and spin degrees of freedom. The {\color{black} SCM}, its' energy and dependence on doping and temperature, as well as the $T$-linear DC resistivity at high $T$ and the upturn at low $T$ can be understood as a {\color{black}charge density wave pinned to impurities, parallel shunted by a metallically conducting component.}

\begin{acknowledgments}

We gratefully acknowledge discussions with C. Berthod and T. Giamarchi. This project was supported by the Swiss National Science Foundation (projects 200021-153405 and 200021-162628.).

K.W. and N.B. contributed equally to this work.
\end{acknowledgments}
\appendix
\section{\color{black}Experimental Methods}\label{app:experiments}
\subsection{Crystal growth and characterization}\label{app:crystal}
\begin{center}
    \begin{table}
        \hfill{}        
        \begin{tabular}{|c|c|}
        \hline
        y (WDS) & $N_e$ (ARPES) \\        
        \hline  
         0    & -     \\  
         0.02 (0.005) & -     \\ 
         0.10 (0.02) & 0.084 (0.030) \\ 
         0.18 (0.04) & 0.13 (0.03)   \\  
        \hline           
        \end{tabular}        
        \hfill{}
        \caption{\label{table:doping} Lanthanum doping in the chemical formula Sr$_{2-y}$La$_y$IrO$_4$ using WDS and number of conduction electrons per formula unit from the Fermi-surface observed with ARPES~\cite{delatorre2015}.}
    \end{table}
\end{center}
\begin{figure}[h!!]
%\centering 
{\color{black}\includegraphics[width=\columnwidth]{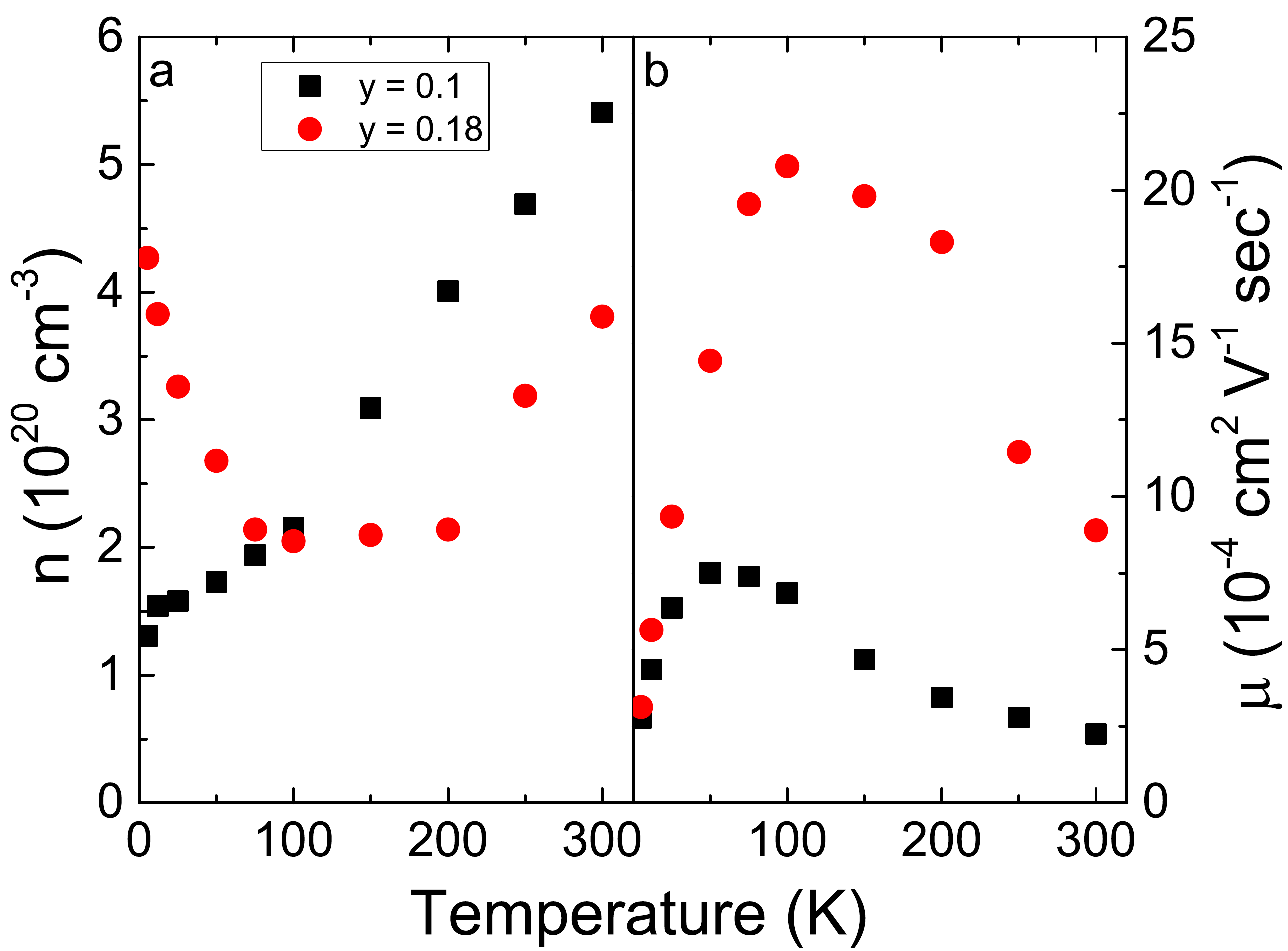}
\caption{\label{fig:Hall}Temperature dependence of Hall density (a) and Hall mobility (b) for the $y=0.1$ and the $y=0.18$ sample.}
}
\end{figure}
Single crystals of Sr$_{2-y}$La$_y$IrO$_4$ were flux grown by heating a mixture of off-stoichiometric quantities of IrO$_2$, La$_2$O$_3$ and SrCO$_3$ in an anhydrous SrCl$_3$ flux to 1245$^\circ$C for 12 hours and cooling the mixture at a rate of $\sim~8^\circ$C/hour to 1100$^\circ$C before quenching to room temperature. 
The typical sample size that was obtained by this method is about 200~$\mu$m to 600~$\mu$m. 
The La concentration was determined by wavelength dispersive spectroscopy (WDS). 
{\color{black}We used the van der Pauw method to measure the DC resistivity shown in Fig.~\ref{fig:resistivity}. For the $y=0.1$ and $y=0.18$ samples we also measured the Hall coefficient, from which we calculated effective carrier density and the mobility, presented in Fig.~\ref{fig:Hall} for the purpose of characterization and comparison to samples used by different research groups. The mobility (righthand panels of Fig.~\ref{fig:Hall}) becomes strongly suppressed when the temperature drops below 50~K (100~K) for the $y=0.1$ ($y=0.18$) sample. The effective densities obtained from the Hall coefficient are smaller than the nominal values La-concentrations ($n=1.03\cdot 10^{21}$ cm$^{-3}$ for the y=0.1 sample and $n=1.85\cdot 10^{21}$ cm$^{-3}$ for the y=0.18 sample). The strong temperature dependence signals that the current is carried by two or more types of charge carriers with different mobilities, and prohibits obtaining the carrier density unambiguously from the Hall data. 
We also measured the Fermi-volume area of samples of the same batch using angle resolved photoelectron spectroscopy (ARPES)~\cite{delatorre2015}  which, by virtue of the Luttinger sum rule, provides the carrier density.} The WDS and ARPES numbers  are summarized in Table~\ref{table:doping}. For the discussion in this paper we will label the samples by the WDS values of $y$ in the first column of this table.
\subsection{Determination of the dielectric function and the optical conductivity.}\label{app:epsilon}
\begin{figure}[h!!]
\includegraphics[width=\columnwidth]{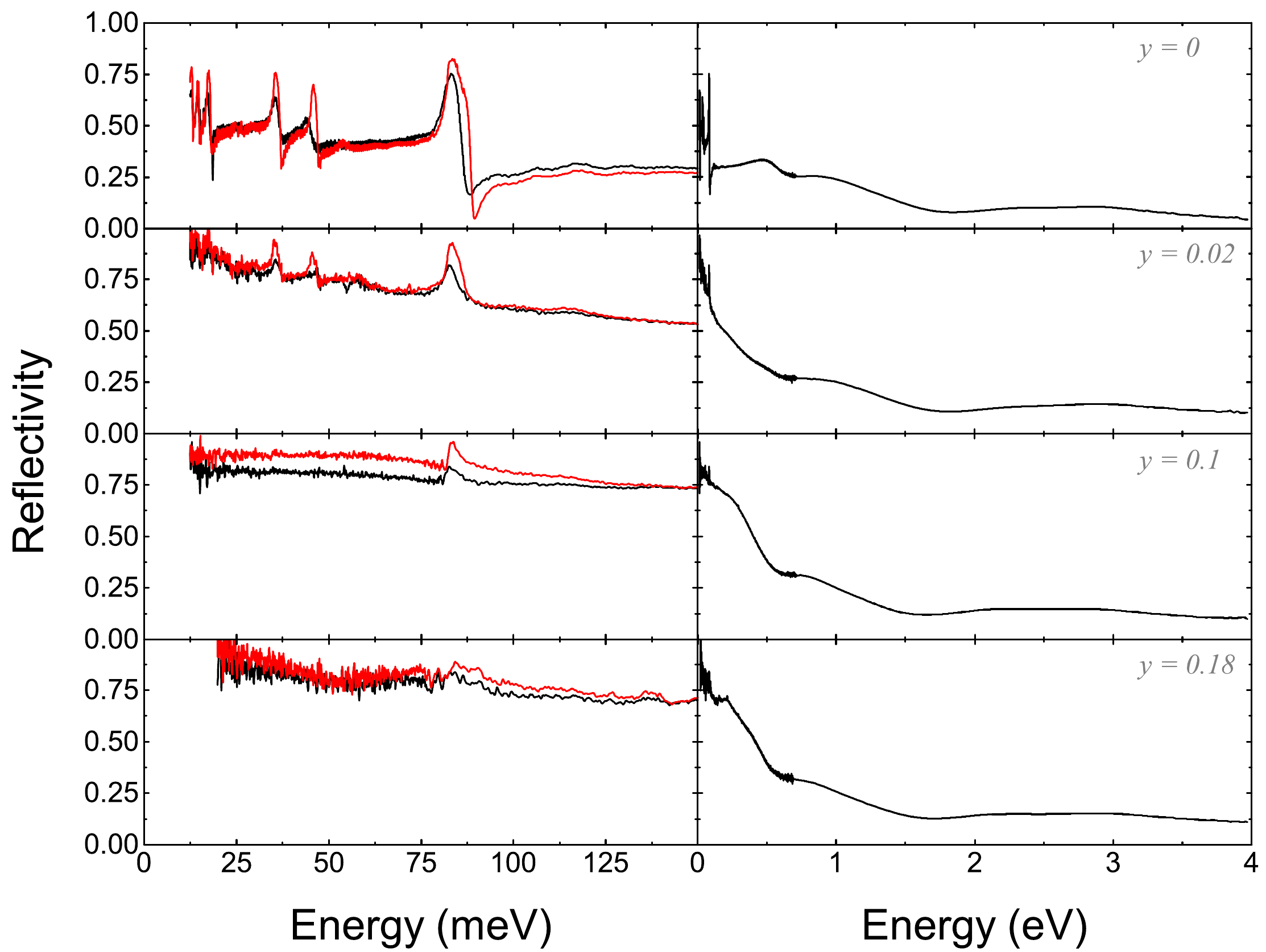}
\caption{\label{fig:reflectivity} Reflectivity spectra at 4~K and 300~K for different doping levels. 
left panel show an expanded scale from 0 to 150~meV, to highlight the range of the optical phonons. 
At 0.7~eV the reflectance and ellipsometry data were merged. 
For the latter only 300 K data could be measured, the whole measured data indicated in right panel. and these data were used for the Kramers-Kronig analysis at all other temperatures using the method detailed in this section.}
\end{figure}
\begin{figure}[h!!]
\includegraphics[width=\columnwidth]{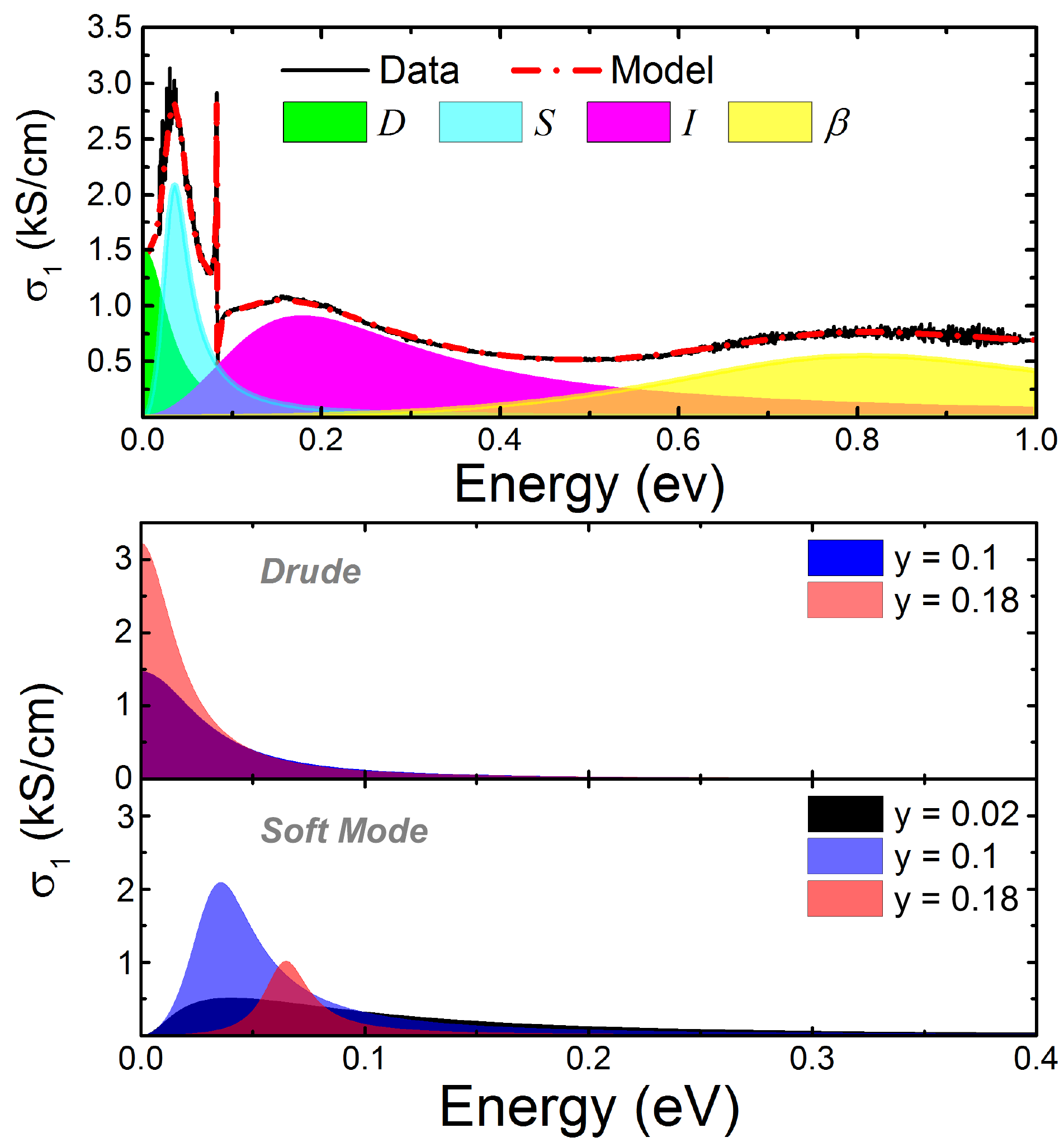}
\caption{\label{fig:DL} Top: The solid  black curve is the real part of the optical conductivity $\sigma_{1}(\omega)$ at 9K below 1eV. 
The Drude-Lorentz fit is indicated by the dashed red line. 
The color coded shaded areas correspond to the Drude (green) and Lorentz (all other colors) oscillators.
Bottom: Drude (top-panel) Lorentz (bottom panel) decomposition below 0.4 eV of the doped samples.} 
\end{figure}
Prior to the optical measurements all samples were cleaved {\em ex-situ}, resulting in clean and mirror-like sample surfaces. We measured the near normal reflectivity from 12.5 meV to 50 meV with a Fourier transform spectrometer combined with a UHV flow cryostat, using in-situ gold evaporation for calibrating the signal. 
The reflectance from 50~meV to 1.25~eV was measured with an infrared microscope and a flow cryostat coupled to a Fourier spectrometer, using for calibration an ex-situ deposited gold layer covering half of the sample surface. 
In the energy range from 0.5~eV to 4~eV we measured the complex dielectric function using  ellipsometry of the ab-plane of our samples at an incident angle 65$^\circ$ relative to the normal. 
Following Aspnes this geometry provides the a-axis tensor element of the dielectric function with a small c-axis contribution~\cite{aspnes1980}. 
Correction for this contribution makes negligible difference in the present case~\cite{propper2016}. 
To obtain sufficient signal-to-noise ratio ellipsometry was performed at room temperature, without cryostat. 
Since the temperature dependence in this energy range is very weak, we use the room temperature data  for the Kramers-Kronig analysis of the infrared data at all temperatures.

The following method was used to obtain the optical conductivity:

(i) From 0.5 to 4~eV the complex $\epsilon(\omega)=\epsilon_1(\omega)+i\epsilon_2(\omega)$ was used to calculate complex reflectivity coefficient $|r|e^{i\phi}$ using Fresnel's equation. 
 The reflectivity spectra $R(\omega)=|r(\omega)|^2$ from 12~meV to 4~eV, combining reflectivity and ellipsometric data, are shown for two temperatures in Fig.~\ref{fig:reflectivity}.

(ii) Fitting the the infrared absolute reflectance $|r(\omega)|$, and visible range complex reflectivity $r(\omega)$ simultaneously with a Drude-Lorentz expansion of $\epsilon(\omega)$ provides extrapolations of $|r(\omega)|$ in the ranges {\color{black}\{0; 12~meV\}} and \{4~eV ; $\infty$\}. 

(iii) Application of the  Kramers-Kronig relation to $|r(\omega)|$ in the range \{0; $\infty$\} provides the phase of the infrared reflectance. 

(iv) Inversion of the Fresnel equation than gives a reliable determination of the complex dielectric function $\epsilon(\omega)$ and the optical conductivity $4\pi\sigma_1(\omega)=\omega\epsilon_2(\omega)$ in the entire range of the experimental data. 

(v) The optical conductivity spectra were binned in 0.3 meV intervals as compared to 0.04 meV of the original reflectivity data shown in Fig.~\ref{fig:reflectivity}.

(vi) In Fig. \ref{fig:rescale} we compare the Kramers-Kronig output without and with an overall offset of the reflectance spectra by $2 \%$. The results demonstrate that the shape and position of the SCM are not significantly affected by this level of uncertainty. {\color{black}Overall, the optical conductivity at low frequencies approaches the zero frequency limit defined by the DC conductivity of this sample.} We have also checked that different extrapolation methods between 0 and 12 meV give the same optical conductivity spectra above 12 meV.
\begin{figure}
\includegraphics[width=\columnwidth]{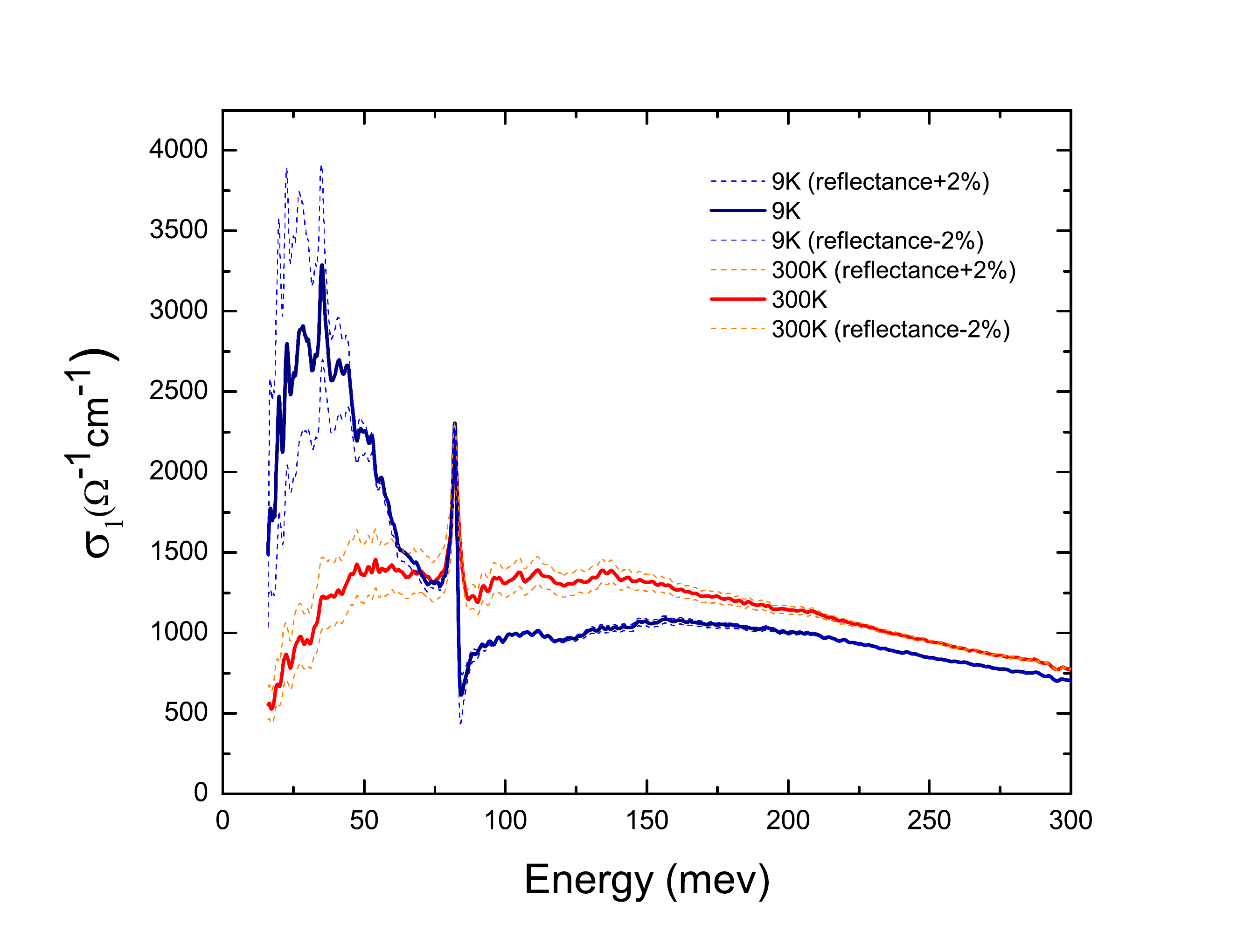}
\caption{\label{fig:rescale} \color{black} Comparison for the $y=0.1$ sample of the Kramers-Kronig output for the optical conductivity with and without an overall vertical shift of the reflectance spectra by $\pm 2 \%$. 
Since our cryostat design with integrated gold evaporator (used for signal calibration) does not involve any mechanical motion of the sample when comparing sample and reference, the systematic error is in fact below $0.5 \%$.
} 
\end{figure}

{\color{black} 
\subsection{Multi-oscillator fit}\label{app:DL}
To characterize the free and bound charge contributions to the optical conductivity, we fitted the experimental data to the following set of expressions for the dielectric function, using the Fresnel equations for the reflectivity and ellipsometry coefficients:
\begin{equation}
\epsilon(\omega)=\epsilon_{\infty}+\frac{4\pi}{i\omega}\sigma(\omega)
\end{equation}
The parameter $\epsilon_{\infty}$ subsumes all bound charge contributions corresponding to interband transitions above the fitted energy range ({\em i.e.} 4 eV), and
\begin{equation}
\sigma(\omega)=
\frac{\omega_p^2}{4\pi}
\frac{\tau}{1-i\omega\tau}
+
\sum_{j}
\frac{S_j\omega_{j}^2}{4\pi}
\frac{\omega+i\omega_j\tan{\theta_j}}{\gamma_{j}\omega+i(\omega^{2}_{j} -\omega^2)}
\end{equation}
Here the first term describes the Drude component, and the sum over $j$ describes all bound charge contributions, including the optical phonons.
The parameter $\theta_j$ describes the Fano-asymmetry of the $j$th optical phonon. All other oscillators could be fitted assuming a Lorentzian profile. The fitting parameters for the phonons of the y=0 sample are reported in Table \ref{table:phonons}.
In Fig.~\ref{fig:DL} the optical conductivity of the 10\% doped sample at 9 K (black curve in the top panel) is shown together with the Drude-Lorentz fit (red dashed line in the top panel). Fits of similar quality were obtained for all other dopings. The middle panel shows the spectral weight of the Drude component for the highest doped samples, and the bottom panel shows the SCM for all doped samples. Note that for the y=0 and y=0.02 samples the Drude weight is negligible, and that there is no SCM for the y=0 sample. }
\begin{center}
    \begin{table}
        \begin{tabular}{|c|c|c|c|c|c|}
        \hline
        $\hbar\omega_j$ & $\hbar\gamma_j$  & $S_j$  & $\theta_j$ & Mode & Type \\
        meV & meV & & degrees & & \\
        \hline  
        12.8	& 0.58 & 1.2 & 0 &E$_u$(1) & External \\ 
        14.3	& 0.61 & 1.2 & 0 &E$_u$(2) & External \\
        17.1	& 0.33 & 1.4 & 0 &E$_u$(3) & External \\
        35.0	& 0.69 & 1.4 & 31 &E$_u$(4) & Bending \\
        45.5	& 0.88 & 0.7 & 42 &E$_u$(5) & Bending \\
        81.8	& 1.8 & 1.0 & 46 &E$_u$(6) & Stretching \\     
        \hline           
        \end{tabular}
        \caption{\label{table:phonons}Fitted parameter values for the optical phonons of the $y=0$ sample.}
      \end{table}
\end{center}
\subsection{AFM and near-field optical microscopy}\label{app:SNOM}
\begin{figure}[h!!]
%\centering 
%\includegraphics[width=\textwidth]{Figure10}
\includegraphics[width=\columnwidth]{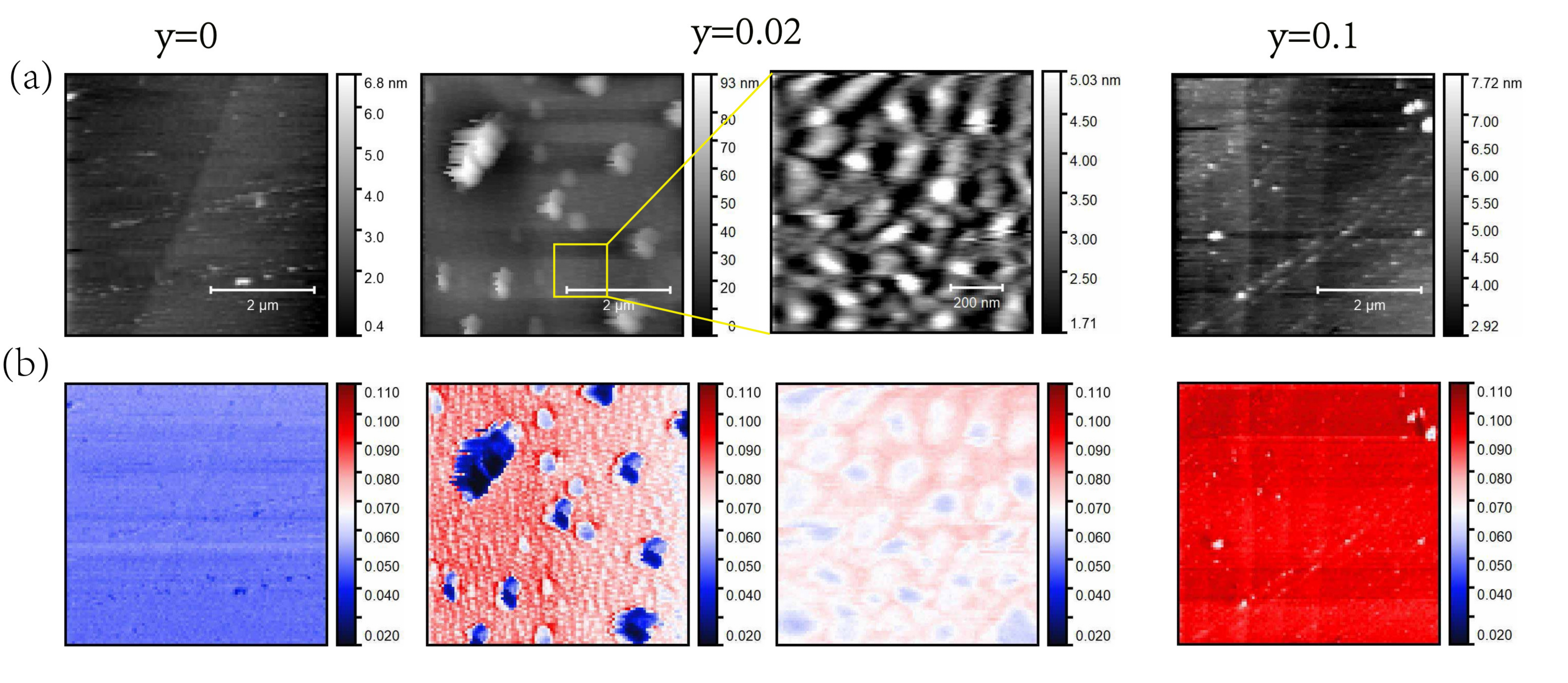}
\caption{\label{fig:SNOM}Atomic force microscopy (a) and near-field infrared optics (b) for a pristine sample and two samples with different doping levels.}
\end{figure}
Scattering-type scanning near-field optical microscopy allows imaging of surface optical properties. Here we used the probing energy of 110~meV ($\sim$ 11$\mu$m wavelength) and the resolution scale is 25~nm. We present images of atomic force topography and the locally back-scattered near-field signal amplitude at room temperature, shown in panels (a) and (b) of Fig.~\ref{fig:SNOM} respectively. Metallic regions where the dc-conductivity is high and the real part of the dielectric function is negative at the probing energy yield high nano-IR signals~\cite{mcleod2016}. The red color represents metallic domains and dark blue represents insulating domains. For the $y=0.02$ doped sample the images demonstrate electronic heterogeneity with insulating islands of 0.2 to 1 $\mu$m size in a metallic background. This corroborates the observation of a remnant of the $\alpha$~peak in Fig.~\ref{fig:sigma_broad} for this doping level. Zooming in on the metallic region displays additional heterogeneity, showing insulating islands on the 50-100 nm scale. The microscopic origin of heterogeneity is most likely originating in local fluctuations of the La spatial distribution, which already present when distributing the La ions randomly, but may be accentuated further due to clustering. These observations confirm the observations with STM on the nanometer scale~\cite{battisti2017}.  A very important point for the present study is the high level of homogeneity of the $y=0.1$ sample on a length scale above 25~nm. This is also the case for the $y=0.18$ sample (not shown here).
\section{Selfconsistent Hartree calculations of the optical properties}\label{app:theory}
We calculated the electronic structure in the subset of $t_{2g}$ states using a tight-binding formalism. We used the following tight-binding Hamiltonian including the Coulomb and exchange interactions within the $t_{2g}$ orbitals ~\cite{marel1988,martins2011,medici2011,arita2012,emori2014,solovyev2015}.
\begin{eqnarray}
\hat{H}&=&\sum_{\langle ij \rangle \alpha \beta \sigma} t_{ij}^{\alpha \beta} \hat{c}^\dagger_{i \alpha \sigma}\hat{c}_{j \beta \sigma}
+\sum_i\left(
\Delta_t  \hat{c}^\dagger_{i d_{xy} \sigma}\hat{c}_{i d_{xy} \sigma}
+\lambda \hat{\vec{L}}_i\cdot \hat{\vec{S}}_i
\right) 
\nonumber\\
&+& U\sum_m\hat{n}_{m\uparrow}\hat{n}_{m\downarrow} 
+\sum_{m>m',\sigma} 
[
(U-2J)\hat{n}_{m',\sigma}\hat{n}_{m,\underline{\sigma}}
\nonumber\\
&+&  
(U-3J)\hat{n}_{m'\sigma}\hat{n}_{m\sigma}
- J  \hat{c}^\dagger_{i m \sigma}
      \hat{c}_{i m \underline{\sigma}}
      \hat{c}^\dagger_{i m'  \underline{\sigma}}
      \hat{c}_{i m' {\sigma}}
\nonumber\\
&-&  J  \hat{c}^\dagger_{i m \sigma}
            \hat{c}^\dagger_{i m \underline{\sigma}}
            \hat{c}_{i m'  \underline{\sigma}}
            \hat{c}_{i m' {\sigma}}
]
\end{eqnarray}
The band structure was calculated treating the interaction terms were treated in the self-consistent Hartree-Fock approximation. The total basis consists of 3 $t_{2g}$ orbitals with two spin-values for each of the two sites. The result is a  $12\times 12$ Hamiltonian, which splits up in two disjoint $6\times 6$ blocks on the bases 
$\{
d_{A,xy,\sigma},
d_{A,yz,\underline{\sigma}},
d_{A,zx,\underline{\sigma}},
d_{B,xy,\sigma},
d_{B,yz,\underline{\sigma}},
d_{B,zx,\underline{\sigma}}
\}$. For each of the two $6\times 6$ blocks one obtains 
\begin{eqnarray}
H(k)=
\left[\begin{array}{llllll}
h_{xy} &\frac{\lambda}{2} &\frac{-i\lambda}{2}&-4 l t_1 &0 &0 \\
\frac{\lambda}{2} &0&\frac{-i\lambda}{2}&0&-2 l t_2 &0\\
\frac{i\lambda}{2}&\frac{i\lambda}{2}&0&0&0&-2 l t_3\\
-4 l^* t_1&0&0&h_{xy}&\frac{\lambda}{2} &\frac{-i\lambda}{2}\\
0&-2 l^* t_2&0&\frac{\lambda}{2} &0&\frac{-i\lambda}{2}\\
0&0&-2 l^* t_3&\frac{i\lambda}{2}&\frac{i\lambda}{2}&0
\end{array}\right]&
\nonumber \\
+
\left[\begin{array}{llllll}
V^H_{A,xy,\sigma}&0 &0 &0 &0&0 \\
0&V^H_{A,yz,\underline{\sigma}}&0 &0 &0&0 \\
0&0&V^H_{A,zx,\underline{\sigma}}&0 &0&0 \\
0&0&0&V^H_{B,xy,\sigma}&0&0 \\
0&0&0&0&V^H_{B,yz,\underline{\sigma}}&0 \\
0&0&0&0&0&V^H_{B,zx,\underline{\sigma}}
\end{array}\right]& 
\nonumber
\end{eqnarray}
where
\begin{eqnarray}
h_{xy}&=&\Delta_t+\epsilon_{xy}\left[\cos(\frac{k_1}{2})\cos(\frac{k_2}{2})\right]^2 \nonumber\\
t_1&=&t_0\cos(\frac{k_1}{2})\cos(\frac{k_2}{2}) \hspace{5mm} 
t_2=t_0\cos\left(\frac{k_1+k_2}{2}\right)\nonumber\\
t_3&=&t_0\cos\left(\frac{k_1-k_2}{2}\right)  \hspace{5mm} 
l=\exp\left(-i\frac{k_1+k_2}{2}\right)\nonumber\\
V^H_{j,xy,\sigma}&=&U\left\langle\hat{n}_{j,xy,\underline{\sigma}}\right\rangle+(U-2J) \left\langle\hat{n}_{j,yz,\underline{\sigma}}+\hat{n}_{j,zx,\underline{\sigma}}\right\rangle \nonumber\\ 
&+& (U-3J) \left\langle\hat{n}_{j,yz,\sigma}+\hat{n}_{j,zx,\sigma}\right\rangle \hspace{3mm}
\mbox{{\em etcycl.}} 
%\nonumber
\end{eqnarray}
Diagonalization of the Hamiltonian is achieved by a unitary transformation
\begin{equation}
\epsilon_{\vec{k},j}=\sum_{\eta,\mu}u^*_{\eta,j}H_{\eta,\mu}u_{\mu,j}(\vec{k})
\end{equation}
The eigenvalues $\epsilon_{k,j}$ and the unitary transformation matrix $u_{\mu,j}(\vec{k})$ were obtained with the lapack routine ZHEEV for diagonalization of Hermitian matrices. 
The optical conductivity was calculated from the expression
\begin{eqnarray}
&{{\mathop{\rm}\nolimits} \mathord{\buildrel{\lower3pt\hbox{$\scriptscriptstyle\leftrightarrow$}} \over \sigma}}
\left(\omega\right)=  
\frac{q_e^2}{\Omega}
\sum\limits_{\vec{k},j}^{{1^t}BZ} 
%\lim_{\vec{k}'\rightarrow\vec{k}}
{\vec{v}} _{j,j}(\vec{k})
{\vec{v}} _{j,j}(\vec{k})
\left(
-\frac{\partial f_{\vec{k},j}}{\partial \epsilon_{\vec{k},j}}
\right)
\frac{i}{\omega  + i\delta } 
+
\nonumber
\\
&+\frac{q_e^2}{\Omega}
\sum\limits_{\vec{k},j\neq m}^{{1^t}BZ} 
%\lim_{\vec{k}'\rightarrow\vec{k}}
{\vec{v}} _{j,m}(\vec{k})
{\vec{v}} _{m,j}(\vec{k})
\frac{f_{\vec{k},j}-f_{\vec{k},m}}
{\varepsilon_{\vec{k},m}-\varepsilon_{\vec{k},j} }
\frac{i\omega}
{\omega \left( \omega  + i\delta\right)-\left( \varepsilon_{\vec{k},m}-\varepsilon_{\vec{k},j} \right)^2} 
\nonumber
\end{eqnarray}
where
\begin{eqnarray}
{\vec{v}} _{j,m}(\vec{k})&=&
\sum\limits_{\eta,\mu}
u_{\eta ,j}^*(\vec{k})u_{\mu ,m}(\vec{k})\nonumber\\
&\times&\left[\frac{\partial}{\partial\vec{k}}{{H}_{\eta ,\mu}(\vec{k})}
+i(\vec{d}_{\mu}-\vec{d}_{\eta}){H}_{\eta ,\mu}(\vec{k})
\right]
\end{eqnarray}
and $\vec{d}_{\mu}$ is the center coordinate of the $\mu$th Wannier orbital in the unit cell.
\begin{center}
    \begin{table}[h!!]
        \begin{tabular}{|c|c|c|c|c|c|c|c|}
        \hline
       \mbox{model}& doping &$t_0$ & $\Delta_t$ & $\epsilon_{xy}$ & $\lambda$ & $U$ & $J$ \\
       && eV & eV &eV & eV &eV & eV \\
       \hline
       A&0&0.23&0.15&-1.5&0.57&2.0&0.0\\
       \hline                 
       A'&0.1&0.35&0.15&-1.5&0.57&0&0\\
       \hline 
       B&y&0.35&0.15&-1.5&0.67&3.1&0.7\\    
       \hline   
       B'&y&0.35&0.15&-1.5&0.67&0&0\\
       \hline              
        \end{tabular}
        \caption{\label{table:theory}Parameter values used for the calculations of the optical conductivity.}
      \end{table}
\end{center}
\begin{figure*}
\includegraphics[width=0.7\textwidth]{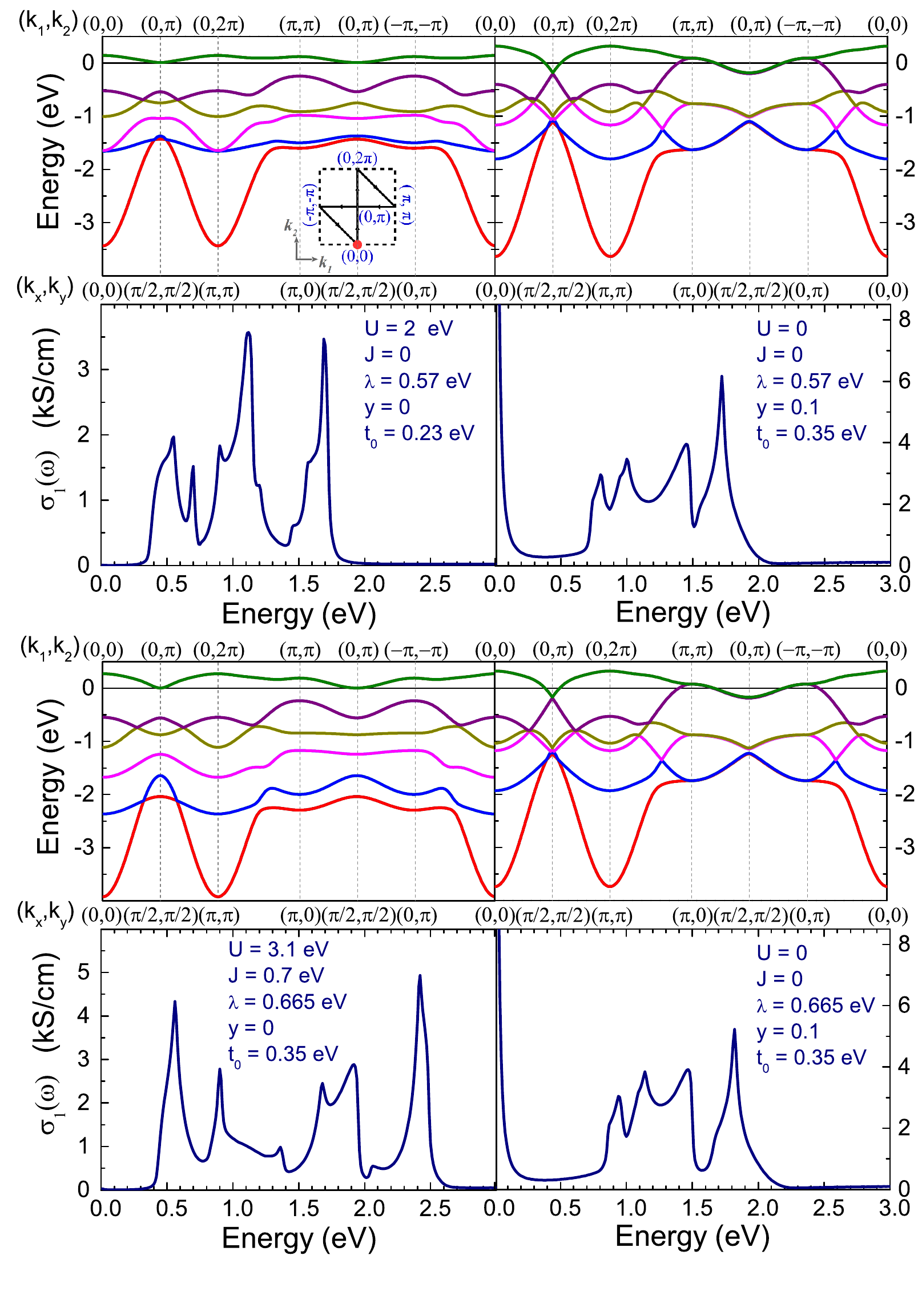}
\caption{\label{fig:theory} 
Top panels:  Band structure (top) and optical conductivity corresponding to model {\color{black}A (left) and A' (right)} of Table~\ref{table:theory}. Different colors are used to distinguish the different bands more easily, but they have no additional meaning.
Bottom panels: Band structure (top) and optical conductivity corresponding to model {\color{black}B (left) and B' (right)} of Table~\ref{table:theory}. Labels $(k_1,k_2))$ along the top refer to the actual $\sqrt{2}\times\sqrt{2}$ unit cell in two dimensions containing 2 Ir atoms. Labels $(k_x,k_y))$  along the bottom refer to the undistorted unit cell in two dimensions containing 1 Ir atom and with $k_x$, $k_y$ along the Ir-O bond direction. 
}
\end{figure*}
{\color{black} Numerical examples are shown in Fig.~\ref{fig:theory} where the Hartree potential resulting from finite $U$ and $J$ was calculated self-consistently. Convergence was reached after about 20 iterations. The parameter values are summarized in Table~\ref{table:theory}.  The parameter sets A (y=0) and A' (y=0.1) were adopted from Ref.~\onlinecite{delatorre2015} for the sake of comparison with the calculations of the energy-momentum dispersion observed with ARPES. The parameter set B  provides the best fit to the experimental data of the undoped parent material (see Fig. \ref{fig:sigma_broad}) and were also used for comparison to the $y=0.18$ doping shown in the same figure, as well as the calculation of $K^*(y)$ in Fig. \ref{fig:spectral_weight}. The parameter set B' was used to calculate $K(y)$ in Fig. \ref{fig:spectral_weight}. }

\par 

From comparing the calculations without and with Hartree potential we see, that the undoped insulator has an indirect gap of about 0.2 eV, with a direct gap at the $(\pi,0)$ point of about 0.6 eV. The direct gap is responsible for the $\alpha$ peak. Whereas the interactions cause serious reshuffling of the optical spectrum, the lowest energy peak in the non-interacting case remains at the same position when $U$ and $J$ are switched on. This energy of this peak closely follows the spin-orbit parameter, and represents transitions from a band with j=3/2  character to the empty  j=1/2 states. Given it's relative robustness we attribute this to the $\beta$ peak in the experimental spectra. We have not been able to adjust parameters such as to fit the entire optical spectrum up to 3 eV. It is conceivable that, as a result of mapping on a limited set of states and in the process of adjusting the low energy part of the spectrum is to experimental data, the dispersion at higher energies gets underestimated.
%
%\newpage
%
\bibliographystyle{apsrev4-1}
%\bibliography{Wang_Manuscript_Bibfile}
%
%
\end{document}